\documentclass[cup6b]{cupbook}
\usepackage{graphicx}
\usepackage{natbib}
\title[Rome, Italy, 27--30 April 2009]
      {The coming of age of X-ray polarimetry}
\author{}
\date{}
\begin{document}
\pagenumbering{arabic}
%%%%%%%%%%%%%%%%%%%%%%%%%%%%%%%%%%%%%%%%%%%%%%%%%%%%

% YOUR CONTRIBUTION GOES HERE

% Insert here the author list and the affiliation
\author[Weisskopf]{Martin C. Weisskopf (NASA/Marshall Space Flight Center) }
% Title of the contribution
\chapter{X-Ray Polarimetry: Historical Remarks and Other Considerations}

% Abstract
\abstract{We briefly discuss the history of X-ray polarimetry for astronomical applications including a guide to the appropriate statistics. We also provide an introduction to some of the new techniques discussed in more detail elsewhere in these proceedings. We conclude our discussion with our concerns over adequate ground calibration, especially with respect to unpolarized beams, and at the system level.}

\section{Introduction\label{s:introduction}}

Sensitive X-ray polarimetry promises to reveal unique and crucial information about physical processes in and structures of neutron stars, black holes, and ultimately all classes of X-ray sources. 
We do not review the astrophysical problems for which X-ray polarization measurements will provide new insights, as these will be discussed in some detail in many of the presentations at this conference.

Despite major progress in X-ray imaging, spectroscopy, and timing, there have been only modest attempts at X-ray polarimetry. 
The last such dedicated experiment, conducted by Bob Novick (Columbia University) over three decades ago, had such limited observing time (and sensitivity) that even $\sim 10\%$ degree of polarization would not have been detected from some of the brightest X-ray sources in the sky. Statistically-significant X-ray polarization was detected in only one X-ray source, the Crab Nebula.

\subsection{History\label{ss:history}}

The first positive detection of X-ray polarization \cite{Netal72} was performed in a sounding rocket experiment that viewed the Crab Nebula in 1971. 
Using the X-ray polarimeter on the Orbiting Solar Observatory (OSO)-8, this result was confirmed \cite{We76} with a 19-$\sigma$ detection ($P =19.2\% \pm 1.0$\%), conclusively proving the synchrotron origin of the X-ray emission. 
Unfortunately, because of low sensitivity, only 99\%-confidence upper limits were found for polarization from other bright X-ray sources (e.g., $\leq 13.5\%$ and $\leq 60\%$ for accreting X-ray pulsars Cen X-3 and Her X-1, respectively \cite{Setal79}.
Since that time, although there have been several missions that had planned to include X-ray polarimeters --- such as the original {\em Einstein} Observatory and Spectrum-X (v1) --- no X-ray polarimeter has actually managed to be launched. 

%\begin{figure}
%\centering
%\includegraphics[scale=0.3]{rocket.eps}
%\caption{1971 Photograph of the NASA Aerobee-350 sounding rocket \#1709 that first detected polarization from the Crab Nebula. Left to right are R. Novick, G. Epstein, M.~C.\ Weisskopf, R. Wolff, \& R. Linke.}
%\label{f:rocket}
%\end{figure}

\section{Instrumental approaches\label{s:instruments}}
There are a limited number of ways to measure linear polarization in the 0.1--50~keV band, sufficiently sensitive for astronomical sources.
We discuss four techniques here, but see also G. Frazier's contribution for a discussion of other techniques.
We emphasize that {\em meaningful X-ray polarimetry is difficult}: 

(i) In general, we do {\em not} expect sources to be strongly ($\gg$10\%) polarized.
For example, the maximum polarization from scattering in an optically-thick, geometrically-thin, accretion disc is only about 10\% at the most favorable (edge-on) viewing angle.
Hence, most of the X rays from such a source carry no polarization information and thus merely increase the background (noise) in the polarization measurement.

(ii) With one notable exception~--- namely, the Bragg-crystal polarimeter --- the modulation of the polarization signal in the detector, the signature of polarization, is much less than 100\% (typically, 20\%--40\%) (and {\em energy-dependent}) even for a 100\%-polarized source.

(iii) The degree of linear polarization is positive definite, so that any polarimeter will always measure a (not necessarily statistically significantly) polarization signal, even from an unpolarized source.
Consequently, the statistical analysis is more unfamiliar to X-Ray astronomers.
For a detailed discussion of polarimeter statistics see \cite{We09}.
The relevant equations are also summarized in slides 18-20 of our presentation.\footnote{http://projects.iasf-roma.inaf.it/xraypol/}.   

Concerning the statistics, one of the most important formulas is the minimum detectable polarization (MDP) at a certain confidence level. In the absence of any instrumental systematic effects, the 99\%-confidence level MDP,

\begin{equation}
{\rm MDP}_{99}=\frac{4.29}{MR_S} [\frac{R_S+R_B}{T}]^{1/2}. \label{e:1}
\end{equation}

where the ``modulation factor'', $M$, is the degree of modulation expected in the absence of background for a $100$\%-polarized beam, $R_S$ and $R_B$ are, respectively, the source and background counting rates, and $T$ is the observing time. 

The MDP is {\em not} the uncertainty in the polarization measurement, but rather the degree of polarization which has, in this case, only a 1\% probablility of being equalled or exceeded by chance. One may form an analogy with the difference between measuring a handful of counts, say 9, with the Chandra X-Ray Observatory and thus having high confidence (many sigmas) that one has detected a source, yet understanding that the value of the flux is still highly uncertain --- 30\% at the 1-sigma level in this example. 
We emphasize this point because the MDP often serves as {\em the} figure of merit for polarimetry.
While it is {\em a} figure of merit that is useful and meaningful, a polarimeter appropriate for attacking astrophysical problems must have an MDP significantly smaller than the degree of polarization to be measured, a point that is often overlooked. 

As P. Kaaret noted during his summary, consider an instrument with no background, a modulation factor of 0.5, and the desire to obtain an MDP of 1\%, this requires detection of $10^6$ counts! 
The statistics will be superb, but the understanding of the response function needs to be compatible. I know of no observatory where the response function is known so well that it may deal with a million count spectrum. 

\subsection{Crystal polarimeters}\label{ss:bragg}

The first successful X-ray polarimeter for astronomical application utilized the polarization dependence of Bragg reflection.
In \cite{We72} we describe the first sounding-rocket experiment using a crystal polarimeter, the use of which Schnopper \& Kalata \cite{SK69} had first suggested.
The principal of operation is summarized in slide 7 of the presentation and a photograph of the one of two crystal panels that focused the X-rays onto a proportional counter was shown in slide 8.

Only three crystal polarimeters (ignoring the crystal spectrometer on Ariel-5 which also served as a polarimeter) have ever been constructed for extra-solar X-ray applications. Only two --- both using graphite crystals without X-ray telescopes --- were ever flown (sounding rocket, \cite{We72}; OSO-8 satellite,\cite{We76}; Spectrum-X (v1) (not flown), \cite{Ketal94}.)

One of the virtues of the crystal polarimeter is, for Bragg angles near 45 degrees, that modulation of the reflected flux approaches 100\%. 
From Eq~\ref{e:1} we see that this is very powerful, {\em all other things being equal}, since the MDP scales directly with the inverse of the modulation factor but only as the square root of the other variables. 
Obviously, a disadvantage is the narrow bandwidth for Bragg reflection.

\subsection{Scattering polarimeters}\label{ss:scattering}

There are two scattering processes from bound electrons --- coherent and incoherent scattering. 
A comprehensive discussion of these processes may be found in many textbooks (see, e.g. \cite{Ja65}).

Various factors dominate the consideration of the design of a scattering polarimeter.
The most important are these: (1) to scatter as large a fraction of the incident flux as possible while avoiding multiple scatterings; (2) to achieve as large a modulation factor as possible; (3) to collect as many of the scattered X-rays as possible; and (4) to minimize the detector background.
The scattering competes with photoelectric absorption in the material, both on the way in and on the way out.
The collection efficiency competes with the desire to minimize the background and to maximize the modulation factor.

Not counting more recent higher energy payloads being developed for balloon and future satellite flights discussed elsewhere in these proceedings, only two polarimeters of this type have ever been constructed for extra-solar X-ray applications. 
The only ones ever flown were suborbital in 1968, see \cite{Aetal69}, in 1969 see \cite{Wetal70}, and in 1971 see, e.g., \cite{Netal72}. The scattering polarimeter \cite{Ketal94} built for the Spectrum-X (v1) satellite was never flown.

The virtue of the scattering polarimeter is that it has reasonable efficiency over a moderately large energy bandwidth, thus facilitating energy-resolved polarimetry.
The principal disadvantage is a modulation factor less than unity, since only for scattering into $90$ degrees will the modulation approach 1.0 in the absence of background, and for a $100$\%-polarized beam. 
To obtain reasonable efficiency requires integrating over a range of scattering angles and realistic modulation factors are under 50\%, unless the device is placed at the focus of a telescope. The modulation factor for the scattering polarimeter on Spectrum-X (v1) reached $\sim$75\%.
At the focus it is feasible to make the scattering volume small which then limits the range of possible scattering angles.

\section{New approaches} \label{s:new_approaches}

In this conference we will hear detailed presentations of a number of new approaches to X-ray (and higher energy) polarimeters. 
We mention two of these approaches here. 

\subsection{Photoelectron tracking} \label{ss:photoelectron}

The angular distribution (e.g., \cite{H54}) of the K-shell photoelectron emitted as a result of the photoelectric absorption process depends upon the polarization of the incident photon. 
The considerations for the design of a polarimeter that exploits this effect are analogous to those for the scattering polarimeter. 
In this case the competing effects are the desire for a high efficiency for converting the incident X-ray flux into photoelectrons and the desire for those photoelectrons to travel large distances before interacting with elements of the absorbing material. 

Here we concentrate on polarimeters that use gas mixtures to convert the incident X-rays to photoelectrons.
Currently there are three approaches to electron tracking polarimetry that use this effect.

To our knowledge, the first electron tracking polarimeter specifically designed to address polarization measurements for X-ray astronomy, and using a gas as the photoelectron-emitting material, was that designed by Austin \& Ramsey at NASA/Marshall Space flight Center (\cite{AR92} - see also \cite{AR93} ; \cite{AMR93}) 
They used the light emitted by the electron avalanches which occur after the release of the initial photoelectron in a parallel plate proportional counter. 
The light was then focused and detected by a CCD camera.

Another gas-detector approach, first discussed by \cite{Cetal01}, uses ``pixillated'' proportional counters (gas electron multipliers) to record the avalanche of secondary electrons that result from gas-multiplication in a high field after drift into a region where this multiplication may take place.
A second approach to such devices was suggested by Black \cite{B07} and exploits time of flight, and rotates the readout plane to be at right angles to the incident flux.
This device sacrifices angular resolution when placed at the focus of a telescope but gains efficiency by providing a greater absorption depth. 

Detecting the direction of the emitted photoelectron (relative to the direction of the incident flux) is not simple because the electrons, when they interact with matter, give up most of their energy at the {\em end} of their track, not the beginning. 
Of course, in the process of giving up energy to the local medium in which the initial photo-ionization took place, the electron changes its trajectory, thus losing the information as to the initial direction and hence polarization. 

\begin{figure}
\centering
\includegraphics[scale=1.0, angle=90.0]{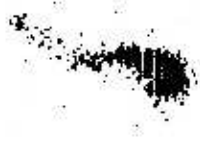}
\caption{The two dimensional projection of a track produced when a 54keV X-ray was absorbed in 2 atm of a mixture of argon (90\%), CH$_4$ (5\%), and trimethylamine (5\%). 
This particular track is $\simeq 14$-mm in length.}
\label{f:etrack}
\end{figure}

It is instructive to examine the image of a track: Figure~\ref{f:etrack} shows one obtained under relatively favorable conditions with an optical imaging chamber.
The initial photoionization has taken place at the small concentration of light to the {\em left} in the figure. 
The size of the leftmost spot also indicates the short track of the Auger electron.
As the primary photoelectron travels through the gas, it either changes direction through elastic scattering and/or both changes direction and loses energy through ionization.
As these interactions occur, the path strays from the direction determined by the incident photon's polarization. 
Of course, the ionization process is energy dependent and most of the electron's energy is lost at the end, not the beginning, of its track.
It should be clear from this discussion that, even under favorable conditions  ---  where the range of the photoelectron is quite large compared to its interaction length --- the ability to determine a precise angular distribution depends upon the capability and sophistication of the track-recognition software, not just the spatial resolution of the detection system.
The burden falls even more heavily on the software at lower energies, where the photoelectron track becomes very short and diffusion in the drifting photoelectron cloud conspires to mask the essential track information.  
Thus, the signal processing algorithms (rarely discussed) form an important part of the experiment, are a possible source of systematic effects, and may themselves reduce the efficiency for detecting polarized X-rays. 

Polarimeters exploiting the photoelectric effect have been discussed in the literature, and two will be discussed in this conference. 
However, {\em no device of this type has ever been flown and those built have undergone relatively limited testing in the laboratory}.
In some cases, performance claims depend more upon Monte-Carlo simulations than actual experiments.
We eagerly await experimental verification of performance at lower energies, around 2-3 keV, where the overall sensitivity peaks.

\subsection{transmission filters}

The potential advent of extremely large-area telescope missions, such as the International X-ray Observatory (IXO), may provide an opportunity to exploit the polarization dependence of narrow-band dichroic transmission filters, as discussed by G. Frazier elsewhere in these proceedings.
The extremely narrow band, a consequent requirement for a detector of a few eV resolution, the low efficiency of the filter, and the association with a major observatory are all issues to be addressed.
Regarding the latter, the history of X-ray polarimetry on major observatories has not been positive.
The OSO-8 polarimeter received only a very limited amount of observing time as the result conflicting pointing requirements.
The Spectrum-X (v1) polarimeter, one of at least two detectors at the focus of its telescope, was allocated only 11 days in the plan for the first year's observing.
The polarimeter on the original Einstein Observatory was ``descoped''.
No polarimeter was selected to be part of either the Chandra or XMM-Newton missions, despite the important capabilities that each of these missions --- especially XMM-Newton with its larger collecting area --- might provide. 

\section{Concluding remarks: systematic effects}

Only a few people in the world have any flight experience with X-ray polarimeters and it behooves one to take advantage of this experience. 
Precision X-ray polarimetry depends crucially on the elimination of potential systematic effects. 
This is especially true for polarimeters with modulation factors less than unity. 
Consider, a polarimeter with a modulation factor of 40\%, and a 5\% polarized source. 
In the absence of any background, this means one is dealing with a signal of only 2\% modulation in the detector. 
To validate a detection means that systematic effects must be understood and calibrated well below the 1\% level, a non-trivial task. 
If present, systematic effects alter the statistics discussed previously, further reducing sensitivity: it is harder to detect two signals at the same frequency than one!. 

To achieve high accuracy requires extremely careful calibration {\em with unpolarized beams}, as a function of energy, at the detector and at the system level! 
For example, suppose that the systematic error in the measured signal of an unpolarized source were 1\%. 
Then for a modulation factor of 40\%, the 3-sigma upper limit due to systematic effects alone would be 7.5\% polarization. 
Thus, if a polarimeter is to measure few-percent polarized sources with acceptable confidence, systematic effects in the modulated signal must be understood at $\leq 0.2\%$.
Careful calibration --- over the full energy range of performance --- is essential.

%%%%%%%%%%%%%%%%%%%%%%%%%%%%%%%%%%%%%%%%%%%%%%%%%%%%
\end{document}